\documentstyle[twoside,aps,prl,multicol,epsf]{revtex} 
\begin{document}
\def\B.#1{{\bbox{#1}}}

\title{{\rm PHYSICAL REVIEW E~~~~{\sl Submitted}\hfill 
 chao-dyn/97010xx  Jan 1997}\\
Perturbative and Non-Perturbative Analysis of the 3'rd Order Zero
Modes \\
in the Kraichnan Model for Turbulent Advection} 
\author { Omri Gat, Victor S. L'vov and Itamar Procaccia} 
\address{Department
of~~Chemical Physics, The Weizmann Institute of Science, Rehovot
76100, Israel} 
\maketitle 
\begin{abstract} 
The anomalous scaling behavior of the $n$-th order correlation
functions ${\cal F}_n$ of the Kraichnan model of turbulent passive
scalar advection is believed to be dominated by the homogeneous
solutions (zero-modes) of the Kraichnan equation $\hat{\cal B}_n{\cal
F}_n=0$. In this paper we present an extensive analysis of the
simplest (non-trivial) case of $n=3$ in the isotropic sector.  The
main parameter of the model, denoted as $\zeta_h$, characterizes the
eddy diffusivity and can take values in the interval $0\le \zeta_h \le
2$. After choosing appropriate variables we can present
computer-assisted non-perturbative calculations of the zero modes in a
projective two dimensional circle. In this presentation it is also
very easy to perform perturbative calculations of the scaling exponent
$\zeta_3$ of the zero modes in the limit $\zeta_h\to 0$, and we
display quantitative agreement with the non-perturbative calculations
in this limit. Another interesting limit is $\zeta_h\to 2$. This
second limit is singular, and calls for a study of a boundary layer
using techniques of singular perturbation theory.  Our analysis of
this limit shows that the scaling exponent $\zeta_3$ vanishes like
$\sqrt{\zeta_2/\log{\zeta_2}}$. In this limit as well, perturbative
calculations are consistent with the non-perturbative calculations.
\end{abstract}
\begin{multicols}{2}
\section{Introduction}
The Kraichnan model of turbulent passive scalar advection\cite{68Kra}
pertains to a field $T(\B.r,t)$ which satisfies the equation of motion
\FL
\begin{equation}
{\partial T(\B.r,t) \over \partial t}+
\B.u(\B.r,t)\cdot\nabla T(\B.r,t)=\kappa \nabla^2
T(\B.r,t)+\xi(\B.r,t). \label{eq}
\end{equation}
Here $\xi(\B.r,t)$ is a Gaussian white random force, $\kappa$ is the
molecular diffusivity and the driving field $\B.u(\B.r,t)$ is chosen
to have Gaussian statistics, and to be ``rapidly varying" in the sense
that its time correlation function is proportional to $\delta(t)$. The
statistical quantities that one is interested in are the many point
correlation functions
\begin{equation}
{\cal F}_{2n}({\bf r}_1,{\bf r}_2,...,{\bf r}_{2n})
 \equiv \langle \!\langle T(\B.r_1,t)
T(\B.r_2,t)\dots T(\B.r_{2n},t)\rangle \! \rangle ,
\end{equation}
where double pointed brackets denote an ensemble average with respect
to the (stationary) statistics of the forcing {\em and } the
statistics of the velocity field.  Assuming that these correlation
functions are scale invariant one is interested in the scaling (or
homogeneity) exponent $\zeta_{2n}$ of ${\cal F}_{2n}$ which is defined
by
\begin{equation}
{\cal F}_{2n}(\lambda \B.r_1,\lambda \B.r_2\dots \lambda
\B.r_{2n}) =\lambda^{\zeta_{2n}} {\cal F}_{2n}(\B.r_1, \B.r_2\dots
\B.r_{2n})\ .
\end{equation}
One expects such a scale invariant solution to exist in the inertial
range, i.e. when all the separations $r_{ij}$ satisfy $\eta\ll
r_{ij}\ll L$ where $\eta$ and $L$ are the inner and outer scales
respectively. It is known \cite{68Kra} that for ${\cal F}_2$ such a
solution exists with $\zeta_2=2-\zeta_h$, where $\zeta_h$ is the
exponent of the eddy-diffusivity, see Eq.~(\ref{eddif}).

The Kraichnan model is unique in the field of turbulence in that it
allows the derivation \cite{94Kra} of an exact differential equation
for this correlation function,
\begin{equation}
\big[- \kappa \sum_{\alpha} \nabla^2_\alpha + \hat {\cal B}_{2n}
 \big] {\cal F}_{2n}({\bf r}_1,{\bf r}_2,...,{\bf r}_{2n})
 = {\rm RHS} . \label{difeq}
\end{equation}
The operator $\hat {\cal B}_{2n}\equiv \sum_{\alpha>\beta}^{2n}
\hat{\cal B} _{\alpha\beta}$, and $\hat{\cal B} _{\alpha\beta}$ are
defined by
\begin{equation}
\hat{\cal B}_{\alpha \beta}\equiv \hat {\cal B}
({\bf r}_\alpha,{\bf r}_\beta)
         = h_{ij}({\bf r}_\alpha -{\bf r}_\beta) \partial^2 /
        \partial r_{\alpha,i} \partial r_{\beta,j}  \ ,
\end{equation}
where the ``eddy-diffusivity" tensor $ h_{ij}(\B.R)$ is given by
\begin{equation}
h_{ij}({\B.R}) = h(R) [(\zeta_h +d-1) \delta_{ij} - \zeta_h R_i
R_j/R^2]\ , \label{eddif}
\end{equation}
and $h(R) = H (R/{\cal L})^{\zeta_h}$. Here ${\cal L}$ is some
characteristic outer scale of the driving velocity field.  The
parameter that can be varied in this model is the scaling exponent
$\zeta_h$; it characterizes the $R$ dependence of $h_{ij}({\B.R})$ and
it can take values in the interval $[0,2]$. Finally, the RHS in
Eq.~(\ref{difeq}) is known explicitly, but is not needed here. The
reason is that it was argued that the solutions of this equation for
$n>1$ are dominated by the homogeneous solutions (``zero-modes"), in
the sense that deep in the inertial interval the inhomogeneous
solutions are negligible compared to the homogeneous one. Also, it was
claimed that in the inertial interval one can neglect the Laplacian
operators in Eq.~(\ref{difeq}), and remain with the simpler
homogeneous equations $\hat {\cal B}_{2n} {\cal F}_{2n}=0$.

Exact solutions of these homogeneous equations are not easy; even in
the simplest case of $n=2$ the function ${\cal F}_{4}$ depends on six
independent variables (for dimensions $d>2$), and one faces a
formidable analytic difficulty for exact solutions. Accordingly,
several groups considered perturbative solutions in some small
parameter, like $\zeta_h$ \cite{95GK} or the inverse dimensionality
$1/d$ \cite{95CFKL}.  The rationale for this approach is that at
$\zeta_h=0$ and $d\to \infty$ one expects ``simple scaling" with
$\zeta_{2n}=n \zeta_2$. The exponent $\zeta_4$, and later also the set
$\zeta_{2n}$, were computed as a function of $\zeta_h$ near these
simple scaling limits. The other limit of $\zeta_h\to 2$ invites
perturbation analysis as well, since one expects that at $\zeta_h=2$
all the scaling exponents $\zeta_{2n}$ would vanish.  Such a
perturbation theory turned out to be elusive.

Recently we reported \cite{us} that it is possible to solve exactly,
eigenfunctions included, the homogeneous equation satisfied by the
3'rd order correlation function ${\cal F}_3(\B.r_1,\B.r_2,\B.r_3)$.
Note that in Kraichnan's model all the odd-order correlation functions
${\cal F}_{2n+1}$ are zero because of symmetry under the
transformation $T\to -T$. This symmetry disappears for example
\cite{example} if the random force $\xi(\B.r,t)$ is not Gaussian (but
$\delta$-correlated in time), and in particular if it has a non-zero
third order correlation
\begin{equation}
{\cal D}_3(\B.r_1,\B.r_2,\B.r_3)\equiv \!\int\!
 dt_1 dt_2\langle \xi(\B.r_1,t_1)
\xi(\B.r_2,t_2)\xi(\B.r_3,0) \rangle. \label{D3}
\end{equation}
With such a forcing the third order correlator is non-zero, and it
satisfies the equation
\begin{equation}
\hat {\cal B}_3 {\cal F}_3(\B.r_1,\B.r_2,\B.r_3) =
 {\cal D}_3\,, \quad
\hat {\cal B}_3\equiv\hat {\cal B}_{12}+
\hat {\cal B}_{13}+\hat {\cal B}_{23}\ . \label{sum}
\end{equation}
This equation pertains to the inertial interval and accordingly we
neglected the Laplacian operators. We also denoted ${\cal D}_3
=\lim_{\B.r_{\alpha\beta} \to 0} {\cal D}_3(\B.r_1,\B.r_2,\B.r_3)$.
The solution of this equation is a sum of inhomogeneous and
homogeneous contributions, and below we study the latter. We will
focus on scale invariant homogeneous solutions which satisfy ${\cal
F}_3(\lambda\B.r_1,\lambda\B.r_2,\lambda\B.r_3)=\lambda^{\zeta_3}
{\cal F}_3(\B.r_1,\B.r_2,\B.r_3)$. We refer to these as the ``zero
modes in the scale invariant sector". We note that the scaling
exponent of the {\em inhomogeneous} scale invariant contribution can
be read directly from power counting in Eq.~(\ref{sum}) (leading to
$\zeta_3=\zeta_2$). Any different scaling exponent can arise only from
homogeneous solutions that do not need to balance the constant RHS.
The scale invariant solutions of Eq.~(\ref{sum}) live in a projective
space whose dimension is lowered by unity compared to the most general
form. These solutions do not depend on three separations but rather on
two dimensionless variables that are identified below. It will be
demonstrated how boundary conditions arise in this space for which the
operator $\hat {\cal B}_3$ is neither positive nor self-adjoint.

In Section~2 we present the transformation of variables that leads to
a precise identification of the projective space. In this space we
present the differential equation that needs to be studied, and derive
the boundary conditions in the projective space. In section~3 we
discuss the perturbation theory that leads to the solution of the
scaling exponents of the zero modes in the limit $\zeta_h\to 0$.  It
is shown that the choice of coordinates of section~2 leads to a
particularly transparent theory in this limit. In section~4 we present
the perturbation theory in the limit $\zeta_h\to 2$. It turns out that
this is a singular perturbation theory, and we discuss the analytic
matchings across boundary layers and near the ``fusion singularity"
which are required to understand this limit, leading to a non-analytic
dependence of $\zeta_3$ on $\zeta_2$.  In Section~5 we deal with the
non-perturbative calculation, culminating with solutions of $\zeta_3$
as a function of $\zeta_h$ throughout the range $0\le \zeta_h \le
2$. It is demonstrated that the non-perturbative solutions are in
agreement with the perturbative calculations at the two ends of this
interval. Section~6 is devoted to a summary and a discussion.
\section{Transformation of Variables}

In this section we describe the transformation of variables in the
operator ${\cal B}_3$ to new variables that are denoted below as
$s,\,\rho,\,\phi$. We first note that equation (\ref{sum}) is
invariant under space translation, under the action of the $d$
dimensional rotation group SO($d$), and under permutations of the
three coordinates. Accordingly, we may seek solutions in the scalar
representation of SO($d$), where the solution depends on the three
separations $r_{12}$, $r_{23}$ and $r_{31}$ only. In the first stage
we transform coordinates to the variables $x_1=|\B.r_2-\B.r_3|^2$,
$x_2=|\B.r_3-\B.r_1|^2$, $x_3=|\B.r_1-\B.r_2|^2$, defining
\begin{equation}
F_3(\B.r_1,\B.r_2,\B.r_3)=f_3(x_1,x_2,x_3) \ .
\end{equation}
By the chain rule,
\begin{equation}
\partial_{1i}F_3(\B.r_1,\B.r_2,\B.r_3)=2(r_{13i}\partial_2f_3+r_{12i}
\partial_3)f_3(x_1,x_2,x_3),
\end{equation}
where $\partial_{1i}\equiv(\partial/\partial_{r_{1i}})$,
 $\partial_2\equiv
(\partial/\partial_{x_{2}})$, and $r_{12i}\equiv r_{1i}-r_{2i}$ etc.

Another application of the chain rule gives
\begin{eqnarray}\label{d1id2j}
\partial_{2j}\partial_{1i}&=&4r_{13i}r_{23j}\partial_1\partial_2+
4r_{12i}r_{23j}\partial_1\partial_3 \\
&& -4r_{13i}r_{12j}\partial_2\partial_3-
4r_{12i}r_{12j}\partial_3^2-2\delta_{ij}\partial_3\ .
\nonumber 
\end{eqnarray}
(For brevity we display only the differential operators explicitly).
Using $2\B.r_{12}\cdot\B.r_{13}=-x_1+x_2+x_3$, and similar identities,
we can now obtain
\begin{eqnarray}
\delta_{ij}\partial_{2j}\partial_{1i}&=&
2(x_1+x_2-x_3)\partial_1\partial_2+
2(-x_1+x_2-x_3)\partial_1\partial_3\nonumber\\&+&
2(x_1-x_2-x_3)\partial_2\partial_3-
4x_3\partial_1\partial_2-2d\partial_3\ , \label{deltaij}
\end{eqnarray}
and
\begin{eqnarray}
{r_{12i}r_{12j}\over r_{12}^2}\partial_{2j}\partial_{1i}&=&
{(x_1-x_2-x_3)(x_1-x_2+x_3)\over x_3}\partial_1\partial_2
\nonumber \\
 +2(-x_1+x_2& -& x_3)\partial_1\partial_3+
2(x_1-x_2-x_3)\partial_2\partial_3\label{rirj}\\
&-& 4x_3\partial_1\partial_2-2\partial_3 \ . \nonumber
\end{eqnarray}
Further calculations, using (\ref{deltaij}) and (\ref{rirj}) give
\begin{eqnarray}\label{B12}
{\cal B}_{12}&\equiv &r_{12}^{\zeta_h}\Big[(d+\zeta_h-1)\delta_{ij}-
\zeta_h{r_{12i}r_{12j}\over r_{12}^2}\Big] 
\partial_{2j}\partial_{1i}
\\ &= &x_3^{\zeta_h/2}\Big [(d-1)o_1+(d-1)(d+\zeta_h)o_2
+{\zeta_h\over x_3}o_3\Big]\ ,
\nonumber 
\end{eqnarray}
where
\begin{eqnarray}
\label{o1}
o_1& = & 2(x_1+x_2-x_3)\partial_1\partial_2+ 2(-x_1+x_2 -x_3)
\partial_1\partial_3\\&& +
2(x_1-x_2-x_3)\partial_2\partial_3-
4x_3\partial_1\partial_2\, ,
\nonumber \\
\label{o2}
o_2& =& -2\partial_3\,,\\
\label{o3}
o_3&=&[x_3(x_1+x_2-x_3)\\
&& -(x_1-x_2-x_3)(x_1-x_2+x_3)] \partial_1\partial_2\ .
\nonumber
\end{eqnarray}
The reader should note that the $o_i$ operators do not depend on the
parameters of the problem. The operators ${\cal B}_{23}$ and ${\cal
B}_{31}$ are obtained from ${\cal B}_{12}$ by cyclic permutations of
the indices, thus completing the transformation of ${\cal B}_3$ to the
$x$ variables.

Note that not every point in the ${x_1,x_2,x_3}$ space corresponds to
a physical configuration. The triangle inequalities between the
pairwise distances translates to the condition
\begin{equation}
\label{ineq}
2(x_1x_2+x_2x_3+x_3x_1)\ge x_1^2+x_2^2+x_3^2.
\end{equation}
This inequality describes a circular cone in the $x_1,\,x_2,\,x_3$
space whose axis is the line $x_1=x_2=x_3$, tangent to the planes
$x_1=0$, $x_2=0$ and $x_3=0$. The
group of permutations between the $x_i$ axes acts very simply on this
cone, corresponding to a $C_6$ operation.

The presence of symmetries motivate a new parameterization of the cone
by three new coordinates $s,\,\rho,\,\phi$:
\begin{eqnarray} \label{trans2}
x_n& =& s [1-\rho\cos( \phi+ \case{2}{3}n \pi) ]\, ,\\
&& 0\le s<\infty,\quad 0\le 
\rho\le1,\quad 0\le\phi\le2\pi \ .\nonumber
\end{eqnarray}
 The new space is a direct product of three intervals, and this fact
 will simplify the discussion of the boundary conditions.  The $s$
 coordinate measures the overall scale of the triangle defined by the
 original $\B.r_i$ coordinates, and configurations of constant $\rho$
 and $\phi$ correspond to similar triangles. The $\rho$ coordinate
 describes the deviation of the triangle from the equilateral
 configuration ($\rho=0$) up to the physical limit of three collinear
 points attained when $\rho=1$; $\phi$ does not have a simple
 geometric meaning. Finally we note that the variables $s,\rho$ and
 $\cos(3\phi)$ are symmetric in the $x_i$ variables. Accordingly any
 function of these variable is automatically invariant under the
 permutation of the $x_i$ variables. We will use this property below.

The final form of the equation is achieved by transforming $o_i$
operators to the variables $s,\rho,\phi$.  To this end we compute the
Jacobian of the transformation (\ref{trans2}) using
 {\sl Mathematica},
\begin{eqnarray}\label{jac}
&&J\equiv{\partial(s,\,\rho,\,\phi)\over
\partial(x_1,x_2,x_3)}= {1\over \sqrt{3} s^2 \rho } \\
&\times& \!\!\! \left[
\begin{array}{ccc}
1,& \! -\rho + \cos (\phi ) + \sqrt{3}\sin (\phi ), & \!
   \sqrt{3}\cos (\phi ) - \sin (\phi ) \\
   1,& \! -\rho + \cos (\phi ) - \sqrt{3}\sin (\phi ),&
  \!  -\sqrt{3}\cos (\phi ) + \sin (\phi ) \\
   1,&\! - \rho -2 \cos (\phi ), &2 \sin (\phi )/\sqrt{3} 
\end{array}\right] .\nonumber
\end{eqnarray}
The Jacobian matrix $J$ is substituted in the chain rule to give the
transformation of the derivatives of $f_3$ with respect to the $x$
variables in terms of the new variables $s,\,\rho,\,\phi$. It is
convenient to perform the tedious calculations using
 {\sl Mathematica} ending up with expressions for 
$o_1,\,o_2,$ and $o_3$.

We present the final long result as a table, in which each item is of
the form 
$$ \{\bar s,\bar
\rho,\bar\phi,\,\,n,\,m,\,c \}
$$ 
representing a term of the form
 $$ c
\rho^n \begin{array}{c}\cos\\ \sin\end{array}(m\phi) s^{\bar s}
\partial_s^{\bar s}
\partial_\rho^{\bar\rho}\partial_\phi^{\bar \phi}.$$
The trigonometric functions cos or sin  appear if $\bar\phi$
 is even or odd respectively.
 The $o_i$ operators are sums of such terms. The tables listing the
 terms appearing in each of the operators are presented in
 Appendix A.

 The upshot of the transformation of the linear operator $\hat {\cal
B}_3$ to the new coordinates is that we derive a second order linear
partial differential operator in the $s,\,\rho,\,\phi$ variables. At
this point we take advantage of the scale invariance of the
differential equation which allows to seek scale invariant solutions
of the form $s^{\zeta_3/2}f(\rho,\phi)$.  Acting on functions of this
form, the operators $s\partial_s$ and $s^2\partial_s^2$ become scalar
multiplications by $\case{1}{2} \zeta_3$ and $\case{1}{2}\zeta_3
(\case{1}{2} \zeta_3-1)$ respectively. The action
of the operator $\hat B_3$ yields an equation for $f(\rho ,\phi)$,
which is the basic equation we study in this paper
\begin{eqnarray}
&&\hat  B_3(\zeta_3) f(\rho,\phi)=[a(\rho,\phi)
\partial_\rho^2+b(\rho,\phi)\partial_\phi^2+
c(\rho,\phi)\partial_\rho\partial_\phi
 \label{rdeq}\\
&&+u(\rho,\phi,\zeta_3)\partial_\rho+v(\rho,\phi,
\zeta_3)\partial_\phi+w(\rho,\phi,\zeta_3)\big ]
f(\rho,\phi)=0 \ .\nonumber
\end{eqnarray}
We note that Eq.~(\ref{rdeq}) can be written in a coordinate-free form
as
\begin{eqnarray}
\label{coofree}
\big [- \B.\nabla\cdot \tensor P(\rho,\phi) \cdot \B.\nabla
& + & \B.q(\rho,\phi, \zeta_3)\cdot\B.\nabla \\
& + & w(\rho,\phi,\zeta_3)]f(\rho,\phi)=0\, ,\nonumber 
\end{eqnarray}
where $\B.\nabla$ is the gradient operator in the $\rho,\phi$ space.
The identification of the tensor $\tensor P$ and the row vector $\B.q$
is obtained by comparison with the explicit form (\ref{rdeq}).  The
new operator $\hat B_3$ depends on $\zeta_3$ as a parameter and it
acts on the unit circle described by the polar $\rho,\phi$
coordinates. The circle represents the projective space of the
physical cone described above. We will see that the availability of a
compact domain (the projective space) will lead to the existence of a
discrete spectrum of the zero modes.

The discrete permutation symmetry of the original Eq.~(\ref{sum})
results in a symmetry of Eq.~(\ref{rdeq}) with respect to the 6
element group generated by the transformation
$\phi\rightarrow\phi+2\pi/3$ (cyclic permutation of the coordinates in
the physical space) and $\phi\rightarrow-\phi$ (exchange of
coordinates). This symmetry extends to a full $U(1)$ symmetry in the
two marginal cases of $\zeta_h=0$ and $\zeta_h=2$ (see
\cite{95SS,96SS} for a discussion of the latter limit) for which all
the coefficients in (\ref{rdeq}) become $\phi$-independent. The
coefficients in (\ref{rdeq}) all have a similar structure, and for
example $a(\rho,\phi)$ reads $$ a(\rho,\phi)=\sum_n[1-\rho
\cos(\phi+\case{2}{3}\pi n)] ^{(\zeta_h-2)/ 2}\tilde
a(\rho,\phi+\case{2}{3}\pi n) \,, $$ where $\tilde a(\rho,\phi)$ is a
low order polynomial in $\rho$, $\cos\phi$ and $\sin\phi$ which
vanishes at $\rho=1,
\phi=0$. We see that the coefficients are analytic
everywhere on the circle except at the three points $\rho=1$,
$\phi=2\pi n/3$ where $n=0,1,2$. These points correspond to the fusion
of one pair of coordinates, and the coefficients exhibit a branch
point singularity there. This singularity is a consequence of the
non-analyticity of the driving velocity field whose eddy-diffusivity is
therefore characterized by a non-integer exponent. This singularity
leads to a nontrivial asymptotic behavior of the solutions which had
been described before in terms of the fusion rules
\cite{96FGLP,96LP}. Since the coefficient $\tilde a$ vanishes at the
fusion point, the singularity is always sub-leading with respect to
terms that come from non-fusing coordinates. Indeed, in
\cite{96FGLP,96LP} it was explained that exposing the singularity
calls for taking a derivative with respect to the fusing
coordinates. Note that for $\zeta_h=2$ the singularity disappears
trivially. For $\zeta_h=0$ there is also no singularity since $\tilde
a$ exactly compensates for the inverse power.

The boundary conditions follow naturally when one realizes that $\hat
B_3$ is elliptic for points strictly inside the physical circle. In
the presentation of Eq.~(\ref{coofree}) this means that
\begin{equation}
\det(\tensor P) > 0 \quad {\rm for}~ \rho<1 .
\end{equation}
This property
is a consequence of the ellipticity of the original operator $\hat
{\cal B}_3$. On the other hand $\hat B_3$ becomes singular on the
boundary $\rho=1$, where the coefficients $a(\rho,\phi)$ and
$c(\rho,\phi)$ vanish. In other words,
\begin{equation}
\tensor P\cdot \B.n|_{\rho=1}=0,
\end{equation}
where $\B.n$ is a unit vector normal to the boundary.  This
singularity reflects the fact that this is the boundary of the
physical region. It follows that $\hat B_3$ restricted to the boundary
becomes a relation between the function $f(\rho{=}1,\phi)\equiv
g(\phi)$ and its normal derivative $\partial_\rho
f(\rho,\phi)\vert_{\rho=1}\equiv h(\phi)$. The relation is
\begin{equation}
bg''(1,\phi)+uh(1,\phi)+vg'(1,\phi)+wg(1,\phi)=0 \ . \label{bc}
\end{equation}
Solutions of Eq.~(\ref{rdeq}) which do not satisfy this boundary
condition are singular, with infinite $\rho$ derivatives at
$\rho=1$. Such solutions are not physical since they involve infinite
correlations between the dissipation (second derivative of the field)
and the field itself when the geometry becomes collinear, but without
fusion.

It is important to stress that the reason for the regularity of the
solutions that do satisfy the boundary conditions is that $\det\tensor
P$ vanishes as a simple zero near the boundary, i.e. like
$(1-\rho)$. Consequently it is possible to find solutions that behave
like $(1-\rho)^0\sim O(1)$ near the boundary.

Finally, we are facing the problem of solving Eq.~(\ref{rdeq})
together with the boundary condition (\ref{bc}). This homogeneous
boundary value problem always have a trivial solution $f=0$. The
condition for the existence of a nontrivial solution is that $\hat
B_3$ is not invertible. We expect that non-invertibility will not be a
generic phenomenon for an arbitrary value of $\zeta_3$. Finding the
set of $\zeta_3$ for which $\hat B_3$ is not invertible becomes a
generalized eigenvalue problem with the exponents $\zeta_3$ playing
the role of eigenvalues. We first discuss this program in the limit of
small $\zeta_h$.
\section{Perturbation Theory near \mbox{$\zeta_{h}=0$}}

The shape of our compact domain invites a Fourier representation in
$\phi$ for the function $f(\rho,\phi)$. The permutation symmetry
implies that only $\cos$ will appear in this representation, and the
index will be divisible by 3. The general equation (\ref{rdeq}) will
then mix different Fourier modes. On the other hand, in the limit
$\zeta_h=0$ the situation simplifies considerably.
\subsection{Zero Modes at $\zeta_h=0$}
The differential equation reduces in the case $\zeta_h=0$ to the
simple form
\begin{equation}\label{zmzh=0}
[\rho^2(1-\rho^2)\partial_\rho^2+\rho(1-d\rho^2)\partial_\rho+
\partial_\phi^2+\lambda\rho^2] f(\rho,\phi)=0,
\end{equation}
where
\begin{equation}
\lambda\equiv\case {1}{2} \zeta_3( \case {1}{2}  \zeta_3+d-1) \ . 
\label{lambda}
\end{equation}
Since the coefficients of Eq.~(\ref{zmzh=0}) are independent of $\phi$
the Fourier modes are decoupled, and we may seek solutions of the form
$f_m(\rho)\cos{m\phi}$ with $m$ divisible by 3. The functions $f_m$
obey the ODEs
\begin{equation}\label{zmode}
[\rho^2(1-\rho^2)\partial_\rho^2+\rho(1-d\rho^2)\partial_\rho
-m^2+\lambda\rho^2] f_m(\rho)=0.
\end{equation} 
Examining the resulting differential equation we note that the
coefficient of the highest derivative in $\rho$ vanishes as a double
zero at $\rho=0$ and as a single zero at $\rho=1$. Since the order of
the zero is not more than the order of the derivative the Frobenius
theory \cite{55CL} of regular singularities is applicable to both
boundaries. Namely, the complete family of solutions in the vicinity
of a given singular point $\rho_0$ is spanned by functions that can be
represented as
\begin{eqnarray}
&&f^{(i)}_m (\rho)=(\rho-\rho_0)^{z_i}\log(\rho-\rho_0)^{k_i}
\sum_{p=0}^\infty a_{i,p} (\rho-\rho_0)^p,
\nonumber\\ &&i=1,2,\quad k_i =0~{\rm or}~1\ , \label{k01}
\end{eqnarray}
where the sum is convergent in a neighborhood of $\rho_0$. One of the
indices $k_i$ is always zero.  When the indices $z_{1,2}$ are
different, we chose arbitrarily $z_1>z_2$ and then $k_1$ is zero.
When $|z_1-z_2|$ is not an integer one also has $k_2=0$. In cases in
which the indices $z_i$ coincide we will chose $k_1=0$, and
$k_2=1$. The numerical values of the indices are obtained by
substituting a solution $(\rho-\rho_0)^{z}$ in the differential
equation, collecting the coefficients of the leading terms near the
singularity, and equating to zero. We refer to the indices $z_i$ as
the ``Frobenius exponents".  Eq. (\ref{zmode}) has regular
singularities at both boundaries $\rho=0,\,1$, with Frobenius exponent
sets of $-m,\,m$ and $0,\,(3-d)/2$ respectively. The singularity at
$\rho=1$ arises from the singularity of the original PDE (\ref{rdeq})
at this boundary, and the boundary condition picks the regular
solution, {\it i.e.,} $f_m(\rho)\sim(1-\rho)^0$ as
$\rho\rightarrow1$. On the other hand, the singularity at $\rho=0$ is
an artifact of the transformation to polar coordinates; however,
analyticity of the solution at $\rho=0$ requires that
$f_m(\rho)\sim\rho^m$ as $\rho\rightarrow0$, specifying the second
boundary condition for (\ref{zmode}).

To proceed with the solution of Eq.~(\ref{zmode}) we make the
transformations $\tilde f_m(\tilde \rho)=\rho^m f_m(\rho)$ and $\tilde
\rho=\rho^2$, and obtain the hypergeometric equation for $\tilde f_m$,
\begin{eqnarray}\label{zmhg}
\big\{ (1& -& \tilde \rho)\tilde \rho\partial_{\tilde \rho}^2+
[ m+1-( m+\case{1}{2}(d+1)] \tilde \rho)\partial_{\tilde \rho} \\
 &&+\case{1}{4}[ \lambda-m(m-1+d)]\big\}
\tilde f_m(\tilde \rho)=0 \ .
\nonumber
\end{eqnarray}
This equation is standard, see \cite{AS}. The unique solution of
Eq.~(\ref{zmhg}) which satisfies the boundary condition at 0 (up to a
multiplication by a constant) is
\begin{equation}\label{hg0}
\tilde f_m(\tilde \rho)=\ _2F_1(a,b;m+1;\tilde \rho),
\end{equation}
where
\begin{equation}\label{hgab}
a=   \case{1}{2}   (m-\case{1}{2}\zeta_3)\, ,
\qquad b=       \case{1}{2}(  m+d-1+ \case{1}{2} \zeta_3)\ .
\end{equation}
It follows from the theory of the hypergeometric functions that the
function $\tilde f_m$ defined in (\ref{hg0}) is regular at $\rho=1$
only if either $a$ or $b$ equal $-n$, where $n$ is a nonnegative
integer. In such a case $\tilde f_m$ becomes a polynomial of degree
$n$. The spectrum of $\zeta_3$ now follows from (\ref{hgab}) and
consists of two sets,
\begin{equation}\label{z3spzh=0}
\zeta_{3(m,n)}^+=2(m+2n),\qquad\zeta_{3(m,n)}^-=-2(d-1+m+2n).
\end{equation}
Since $\lambda(\zeta_{3(m,n)}^+)=\lambda(\zeta_{3(m,n)}^-)$,
[cf. Eq.~(\ref{lambda})], the corresponding eigenfunctions depend only
on $n$ and $m$. Expressed in terms of the original variables the
eigenfunctions are
\begin{eqnarray}
\label{hgef} &&f_{(m,n)}(\rho,\phi)\\&&=
\rho^m\ _2F_1(-n,n+m+(d-1)/2;m+1;\rho^2)\cos(m\phi) \ . 
\nonumber
\end{eqnarray}

The zero-modes of the first set, with positive values of $\zeta_3$ are
the ones to be matched at the outer scale. Of these, the most relevant
non-trivial zero mode is $f_{0,1}$ with $\zeta_{3(0,1)}^+=4$, still
less relevant than the scaling of the forced solution.
\subsection{Zero Modes for {$0<\zeta_h\ll 1$}}
Perturbation theory for $\zeta_h\rightarrow0$ will be carried out by
expanding the solution for small $\zeta_h$ in terms of the $\zeta_h=0$
eigenfunctions $f_{(m,n)}$, in a procedure very similar to time
independent perturbation theory in quantum mechanics. We first cast
Eq. (\ref{zmode}) in Sturm-Liouville form
\begin{eqnarray}\label{zmsl}
\Big \{{1\over \rho(1-\rho^2)^{(d-3)/2}}{\partial_\rho}\Big[
\rho(1&-& \rho^2)^{(d-1)/2}
\partial_\rho\Big] \\ 
& - & {m^2\over\rho^2}\Big\} f_m(\rho)
= -\lambda f_m(\rho)\  .\nonumber
\end{eqnarray}
It follows that the zero-modes $f_{(m,n)}$ form an orthogonal set with
respect to the inner product
\begin{equation}\label{zmip}
\left<f,g\right>_0\equiv\int\limits_{\rho\le1} {
d\rho d\phi\over2\pi}
 \rho(1-\rho^2)^{(d-3)/2}f (\rho,\phi)g(\rho,\phi)\ .
\end{equation}
We now assume that we may expand the eigenfunction and eigenvalues as
\begin{eqnarray}
f(\rho,\phi)& = & f_{(m,n)}(\rho,\phi)+\zeta_h f^{(1)}(\rho,\phi)+
O(\zeta_h^2)\, ,\\
\zeta_3& = & \zeta_{3(m,n)}+\zeta_3^{(1)}+O(\zeta_h^2)\, ;
\end{eqnarray}
writing
\begin{eqnarray} 
B(\zeta_h,\zeta_3)= B(0,\zeta_{3(m,n)})& +&
\zeta_h\Big[\partial_{\zeta_h}B(0,\zeta_{3(m,n)})\\
&&+\zeta_3^{(1)}\partial_{\zeta_3}B(0,
\zeta_{3(m,n)})\Big]+O(\zeta_h^2)\, , \nonumber
\end{eqnarray}
Eq (\ref{rdeq}) becomes, to order $\zeta_h$,
\begin{eqnarray} \label{zmsmallzh}
B(0,\zeta_{3(m,n)})f^{(1)}(\rho,\phi) +
\Big[\partial_{\zeta_h}B(0,\zeta_{3(m,n)}) && \\
+\zeta_3^{(1)}\partial
_{\zeta_3}B(0,\zeta_{3(m,n)})\Big]f_{(m,n)}
(\rho,\phi)&=&0 \ .\nonumber
\end{eqnarray}
Taking the inner product $\left<\right>_0$ of Eq. (\ref{zmsmallzh})
with $f_{(m,n)}$, and using the self-adjointness of
$B(0,\zeta_{3(m,n)})$ with respect to this inner product, there
follows an expression for the first correction to $\zeta_3$,
\begin{equation}\label{z31}
\zeta_3^{(1)}=-{
\left<f_{(m,n)},\partial_{\zeta_h}B(0,\zeta_{3(m,n)})
f_{(m,n)}\right>_0\over
\left<f_{(m,n)},\partial_{\zeta_3}B(0,\zeta_{3(m,n)})f_{(m,n)}
\right>_0}\ .
\end{equation}
The integrals in (\ref{z31}) were programmed using {\sl Mathematica},
and we used the program to generate expressions for the first few
values for $\zeta_3^{(1)}$ presented below:
\end{multicols}
\renewcommand{\arraystretch}{1.2}
\begin{equation}\begin{array}{rrcclc}
m&n&\zeta_{3(m,n)}^+&\zeta_3^{+(1)}&\zeta_{3(m,n)}^-&\zeta_3^{-(1)}\\
0&0&0&0&-2d+2&-1\\
0&1&4&{2(2-d)\over d-1}&-2d-2&3-d\over d-1\\
0&2&8&{{2\,\left( 159 + 91\,d - 31\,{d^2} 
- 19\,{d^3} - 2\,{d^4} \right) }\over
   {-48 + 4\,d + 32\,{d^2} + 11\,{d^3} + {d^4}}}&-2d-6&
{{-3\,\left( 90 + 62\,d - 10\,{d^2} - 9\,{d^3} - {d^4} \right)
}\over {-48 + 4\,d + 32\,{d^2} + 11\,{d^3} + {d^4}}}\\
0&3&12&{{6\,\left( 780 + 503\,d - 89\,{d^2}
 - 103\,{d^3} - 19\,{d^4} -{d^5}
\right)}\over {-480 - 8\,d + 324\,{d^2} + 142\,{d^3} + 21\,{d^4} +
{d^5}}}&-2d-10&-{{4200 + 3010\,d - 210\,{d^2} 
- 476\,{d^3} - 93\,{d^4} -
      5\,{d^5}}\over  {-480 - 8\,d + 324\,{d^2} 
+ 142\,{d^3} + 21\,}}\\
3&0&6&{{-3\,\left( 15 + 8\,d + {d^2} \right) }\over {-2 + d +{d^2}}}
&-2d-4& -{{\left( 5 + d \right) \,\left( 7 + 2\,d \right) }\over
    {2 - d - {d^2}}}\\
3&1&10&-{{\left( 9 + d \right) \,\left( -489 - 131\,d + 136\,{d^2} +
        53\,{d^3} + 5\,{d^4} \right) }\over
    {-144 - 36\,d + 100\,{d^2} + 65\,{d^3} + 14\,{d^4} + {d^5}}}
&-2d-8&
-{{\left( 9 + d \right) \,\left( 441 + 135\,d - 104\,{d^2} -
        42\,{d^3} - 4\,{d^4} \right) }\over
    {-144 - 36\,d + 100\,{d^2} + 65\,{d^3} + 14\,{d^4} + {d^5}}} \\
6&0&12&
{{3\,\left( 3 + d \right) \,{{\left( 11 + d \right) }^2}\,
     \left( 370 + 216\,d + 37\,{d^2} + 2\,{d^3} \right) }\over
   {3840 + 544\,d - 2584\,{d^2} - 1460\,{d^3} - 310\,{d^4} -
     29\,{d^5} - {d^6}}}&-2d-10&
{{5\,\left( 11 + d \right) \,
     \left( 6558 + 6508\,d + 2409\,{d^2} + 412\,{d^3} +
       33\,{d^4} + {d^5} \right) }\over
   {-3840 - 544\,d + 2584\,{d^2} + 1460\,{d^3} + 310\,{d^4} +
     29\,{d^5} + {d^6}}}
\end{array}
\end{equation}
\begin{multicols}{2}
\section{Perturbation Theory in the limit $\zeta_h\to 2$}
In this section we treat the case $\zeta_2=2-\zeta_h\ll 1$
perturbatively. The perturbation theory is singular, since the leading
order approximation, obtained by setting the small parameter
$\zeta_2=0$ is not valid throughout the entire domain. We employ
boundary layer techniques and matching of asymptotics to obtain an
asymptotic approximation for $\zeta_3$ in this limit. As expected,
$\zeta_3$ goes to 0 with $\zeta_2$, but with a non-trivial dependence
\begin{equation}\label{mag}
 \zeta_3=O\big(\sqrt{ \zeta_2/\log\zeta_2}\big)\ .
\end{equation}
\subsection{Leading order solution}
As in the case $\zeta_h=0$ discussed above, substituting $\zeta_h=2$
into the zero-mode equation (\ref{rdeq}) yields an equation with
coefficients which are independent of the variable $\phi$.
This equation  reads
\begin{eqnarray}
&&[\rho^2(1-\rho^2)^2\partial_\rho^2
 + \rho(1-\rho^2)(1-2\rho^2+\zeta_3\rho^2)
\partial_\rho  \\
&& + (1-\rho^2)\partial_\phi^2
+w(\rho)]f(\rho,\phi)=0  \,, \nonumber \\
&&w(\rho)=\rho^2{\zeta_3\over 2d}\Big[(d+2)(d-1)+({\zeta_3\over 2}
-1)\Big((d-1)(\rho^2+1)\nonumber\\&&+\rho^2-1\Big)\Big] \ . 
\label{eqz=2}
\end{eqnarray}
This significant simplification is a consequence of the higher
symmetry of the passive scalar equation in this limit, as discussed in
detail in Ref~.\cite{95SS,96SS}.

Making use of the symmetry, we look for solutions of the form
\begin{equation}f(\rho,\phi)=f_m(\rho)\cos m\phi \quad m=3n,\,n=0,1,2
\dots
\end{equation}
The functions $f_m$ satisfy the ODEs
\begin{eqnarray}
&&[\rho^2(1-\rho^2)^2\partial_\rho^2+\rho(1-\rho^2)(1-2\rho^2
+\zeta_3\rho^2)
\partial_\rho\nonumber\\&&-m^2(1-\rho^2)+w(\rho)]f_m(\rho)=0
 \ . \label{z2ode}
\end{eqnarray}
Equation (\ref{z2ode}) has regular singularities at the points 0
and 1. In close analogy with the case
$\zeta_h=0$ we find that the Frobenius exponents at 0 are $\pm m$, and
choose the solution which behave as $\rho^m$ at 0, thus providing the
boundary condition at 0.

The behavior near the boundary $\rho=1$ is different. In contrast with
the case $0\le\zeta_h<2$ where one of the Frobenius exponents is
always 0, when $\zeta_h=2$ the Frobenius exponents depend on the value
of $\zeta_3$, and generally neither is 0. Thus, the qualitative
behavior of the zero modes for $0<\zeta_2=2-\zeta_h\ll 1$ near
$\rho=1$ is different from that of the solutions of Eq.~(\ref{eqz=2})
where $\zeta_h$ is set equal to 2. In other words, we expect the
existence of a boundary layer near $\rho=1$; Eq.~(\ref{eqz=2})
describes well only the behavior of the {\it outer} solution, namely
the leading order approximation away from the boundary layer.

The outer solution may now be written explicitly. Using standard
transformations \cite{AS} Eq.~(\ref{z2ode}) may be reduced to a
hypergeometric equation, whose solution which obeys the $\rho=0$
boundary condition is
\begin{equation}
f_m(\rho)=\rho^m~_2F_1(m/2,(m+1)/2;m+1;\rho^2)
\ . \label{out}
\end{equation}
Since we anticipate that $\zeta_3\ll 1$, we give in (\ref{out}) the
solution of equation (\ref{z2ode}) taking $\zeta_3=0$.  This
approximation is justified to leading order {\it a-posteriori}, when we
find that $\zeta_3$ is indeed small. The outer solution is constructed
from the $f_m$'s by
\begin{equation}\label{outer}
f^{out}(\rho,\phi)=\sum_{m=0,3,6,\dots} \nu_m f_m(\rho) \cos{m\phi}
\end{equation}
The coefficient $\nu_m$ are unknown at this stage and they will be
determined by matching with the inner solution.

The outer solution is a valid asymptotic approximation of the true
solution only when $\zeta_2\ll1-\rho$.  Since we expect the solution
to vary very rapidly within a boundary layer near $\rho=1$, the inner
solution, valid within the boundary layer, should be expressed in
terms of a ``fast'' variable $\tau$ that changes on a scale inversely
proportional to $\zeta_2$, namely,
\begin{equation}
\rho=1+\zeta_2 \tau.
\end{equation}
We therefore change the variable $\rho$ to $\tau$ in the differential
equation (\ref{rdeq}) and keep
the leading terms in $\zeta_2$. The resulting equation for the inner
solution is
\begin{eqnarray}
&&\{\tau[\tau-{d+1\over 2d}p(\phi)]\partial^2_\tau +
({(d+3)(d-1)\over 4d}p(\phi)-{1\over2}\tau)
\partial_\tau\nonumber\\&&+w(1)/4\}f^{\rm in}(\tau,\phi)=0 \ ,
 \label{eqin}
\end{eqnarray}
where
\begin{equation}\label{pphi} p(\phi)=\sum_{\phi'=\phi,\,\phi\pm
2\pi/3}\log(1-\cos\phi')(\cos{\phi'}-\cos{2\phi'}), 
\end{equation}
and the function $w$ is defined in (\ref{eqz=2}). It is crucial for
the matching between inner and outer solution to realize that
Eq.~(\ref{eqin}) is not valid near the fusion points $\phi=2\pi n/3$.
The significance of this fact is explained below.

Equation~(\ref{eqin}) has a regular singularity at $\tau=0$ which
corresponds to the $\rho=1$ boundary, and as in the other cases
demanding regularity gives a boundary condition at this point.  The
equation can be again transformed into the hypergeometric equations, so
it has solutions of the form
\begin{eqnarray}
&&f^{\rm in}(\tau,\phi)=\mu(\phi)\label{in}\\&&\times
~_2F_1\left(-{(d-1)\zeta_3\over 4},
-{1\over2};{(d+3)(d-1)\over2(d+1)};{2d\over d+1}{\tau\over p(\phi)}
\right)
\ . \nonumber
\end{eqnarray}
Here again we used the smallness of $\zeta_3$ and neglected terms of
higher order in it. The function $f^{\rm in}$ provides an asymptotic
approximation to the actual solution when $-\tau\ll(1/\zeta_2)$. It
depends on the function $\mu(\phi)$ that will be determined by
matching to the outer solution.

\subsection{Asymptotic matching}
The next step involves matching of the two approximations $f^{\rm in}$
and $f^{\rm out}$ in their common region of validity
\begin{equation}
\zeta_2 \ll 1-\rho \ll 1 \ . \label{cond}
\end{equation}
We perform the matching using standard boundary layer methods by
examining the asymptotic behavior of $f^{\rm in}$ and $f^{\rm out}$ in
the matching region (\ref{cond}) and balancing coefficients. We make
use of the asymptotic behavior of the hypergeometric functions
\cite{AS}. For the outer solution we need the asymptotics of the
hypergeometric function with argument close to 1, yielding
\begin{equation}\label{asout}
f^{\rm out}(\rho,\phi) \begin{array}{c}~\\
\!\sim\!
\\ \scriptstyle1-\rho\ll 1\end{array}
\sum_m2^m\nu_m\Big [1\!-m \sqrt{2(1-\rho) }\Big]
\!\cos(m\phi)\ .
\end{equation}
For the inner solution we use the behavior of the hypergeometric
function for large negative values, and obtain, using the relation
between $\rho$ and $\tau$
\begin{equation}\label{asin}
f^{\rm in}(\rho,\phi) \begin{array}{c}~\\ 
\! \sim \!
\\ \scriptstyle1-\rho\ll 1\end{array}
\mu(\phi)\Bigg [1+ c\zeta_3  
\sqrt{1-\rho \over \zeta_2 p(\phi)}\ \Bigg  ]\, ,
\end{equation}
where
\begin{equation}
c={ \sqrt{\pi}\, \Gamma\big[{(d+3)(d-1)\over2(d+1)}\big]
\over\Gamma\big[{(d+3)(d-1)\over2(d+1)}
+{1\over2}\big]}{d-1\over4}\sqrt{d+1\over 2d}.
\end{equation}

To proceed with the matching of (\ref{asout}) and (\ref{asin}), we
first match the $O(1)$ terms [for small $(1-\rho)$], giving
\begin{equation}
\label{o(1)}\mu(\phi)=\sum_m 2^m\nu_m\cos(m\phi),\end{equation}
{\it i.e.} $2^m\nu_m$ are just the coefficients of the Fourier series
of $\mu(\phi)$. Next we have to match the $O(\sqrt{1-\rho}\, )$ terms
giving
\begin{equation}\label{o(2)} 
{c\zeta_3\over \sqrt{\zeta_2  p(\phi)}}\mu(\phi)=
\sum_m -m\sqrt{2}~2^m\nu_m\cos(m\phi)\ . 
\end{equation}
We expand $p(\phi)^{-1/2}$ in Fourier series,
\begin{equation} p(\phi)^{-1/2}=\sum_m p_m \cos(m\phi)
 \end{equation}
and substitute in (\ref{o(2)}) to get, using (\ref{o(1)})
\begin{eqnarray}
&&\sum_{mn} {c\zeta_3\over 2\sqrt{\zeta_2}}p_n\nu_m
(\cos[(n-m)\phi]+\cos[(n+m)\phi])                   \\
&=&
\sum_m -m\sqrt{2}\nu_m\cos(m\phi)\ . \nonumber
\end{eqnarray}
Equating Fourier coefficients of the same order yields finally
\begin{equation}\label{match}
\sum_n{c\zeta_3\over 2\sqrt{\zeta_2}}p_n(\nu_{m+n}+
\nu_{|m-n|})=-m \nu_m. 
\end{equation}
The matching condition (\ref{match}) is a generalized eigenproblem for
the infinite vector $\nu_m$, with eigenvalues $\zeta_3$.

Since $\zeta_3$ appears in (\ref{match}) only through the combination
$\zeta_3/\sqrt{\zeta_2}$, one would be led to conclude that
$\zeta_3=O(\sqrt{\zeta_2}\, )$ for all non-trivial solutions of
(\ref{match}) with different numerical coefficients. This
consideration is modified, however, since the coefficients $p_n$
themselves also depend on $\zeta_2$ as will be now demonstrated.

It follows from the definition of $p(\phi)$ (\ref{pphi}), that
\begin{equation}
p(\phi)=O[\phi^2\log(\phi)]
\end{equation}
for $\phi$ small. Note that the Fourier coefficients $p_n$ are written
as integrals
\begin{equation}
p_n={1\over \pi}\int_0^{2\pi} {d\phi\over \sqrt{p(\phi)}}\quad
\mbox{for}\ \  n>1 \ .
\label{pm}
\end{equation}
All these integrals
diverge at $\phi=0$, which is precisely the fusion point where the
approximation (\ref{in}) ceases to be valid, requiring a more careful
examination of the behavior in this region. In deriving Eq.~(\ref{in})
we made the approximation
\begin{eqnarray}
(1-\rho\cos\phi)^{-\zeta_2/2}&\sim& 1-\case{1}{2} \zeta_2
\log(1-\rho\cos\phi)\nonumber\\&\sim& 1-\case{1}{2}\zeta_2
\log(1-\cos\phi)\ ,
\end{eqnarray}
for $1-\rho=O(\zeta_2)$. When $\phi^2=O(\zeta_2)$ the second
approximation is no longer valid, and one has instead
\begin{equation} (1-\rho\cos\phi)^{-\zeta_2/2}
\sim 1-\case{1}{2}\zeta_2\log[(1-\rho)+\case{1}{2}\phi^2]\,,
\end{equation}
so that instead of (\ref{eqin}) we get a similar equation in which
$p(\phi)$ is replaced by $\log(-\zeta_2\tau+\case{1}{2}\phi^2)$. The
resulting equation is no longer integrable in terms of hypergeometric
functions.  However, we still expect the asymptotic approximation
(\ref{asin}) to be valid, but for small $\phi$ the function $p(\phi)$
is no longer
given by (\ref{pphi}). We may estimate $p(\phi)$ for small $\phi$ by
the following consideration. Examining Eq. (\ref{eqin}) we see that
$p(\phi)$ is the value of $\tau$ where the coefficient of the second
derivative crosses over from quadratic behavior to linear behavior,
and the coefficient of the first derivative crosses over from linear
to constant behavior. When $\phi^2=O(\zeta_2)$ there is also a
crossover, but to a linear or constant function times a logarithmic
function of $\tau$. The logarithmic function changes very slowly, and
may be approximated roughly by a constant. The crossover occurs when
\begin{equation}
|\tau|\sim\zeta_2\log(\zeta_2|\tau|+\phi^2)
\sim\zeta_2\log(\zeta_2)\ . 
\end{equation}

We are thus led to make the following approximation
\begin{equation}p(\phi)\sim \zeta_2\log\zeta_2,\quad{\rm for}\quad
\phi^2=O(\zeta_2)\ . 
\end{equation}
This estimate implies that when calculating the Fourier coefficients
in (\ref{pm}) the integral should be cut off at $\phi\sim
\sqrt{\zeta_2}$, giving the coefficients a $\case{1}{2}\log(\zeta_2)$
 dependence on $\zeta_2$.  It is now possible to balance powers in the
 matching Eq.~(\ref{match}), obtaining the order of magnitude
 relation~(\ref{mag}).

\section{Solutions for General Values of $\zeta_h$}
For general values of $\zeta_h$ and $d$ the differential
equation~(\ref{rdeq}), which has variable coefficients is not
accessible to analytic techniques.  In this section we present
numerical solutions of the scaling exponents $\zeta_3$ for arbitrary
values of $\zeta_h$, using a discretized version of the operator $\hat
B_3$. Since the differential problem is a linear homogeneous equation
with linear homogeneous boundary conditions, the discretized problem
is also a homogeneous linear equation, implying that non-trivial
solutions exist only when the determinant of the discretized operator
vanishes. This determinant depends parametrically on $\zeta_3$.  Since
the differential operator is defined on a compact domain we expect the
determinant to vanish only 
\narrowtext
\begin{figure}
\epsfxsize=8truecm
\label{Fig1}
\caption{
 The scaling exponent $\zeta_3$ as a functions of $\zeta_h$ found as
 the loci of zeros of the determinant of the matrix , for d=2. }
\end{figure}
\noindent
for discrete values of $\zeta_3$ for any given values of $\zeta_h$ and
the dimensionality $d$. One solution is known to exist always, a
constant zero-mode associated with $\zeta_3=0$. Our aim is to find the
lowest lying positive real solutions $\zeta_3$ for which the
determinant vanishes.

The discretization of the operator $\hat B_3$ was carried out as
follows: We defined a nine-point finite difference scheme for the
second order equation (\ref{rdeq}). The discretization of the boundary
conditions at $\rho=1$ (\ref{bc}) is achieved using the same scheme,
which requires in this case only three boundary points and one
interior point, since on the boundary the radial derivative appears in
first order.  Using the symmetry of the problem we restricted the
domain to one sixth of the circle, $0<\phi<\pi/3$.  The symmetry
implies that original problem on the full circle is equivalent to the
problem on the reduced domain with simple Neuman boundary conditions
$\partial_\phi f(\rho,\phi)=0$.  on the new boundary lines
$\phi=0,\pi/3$.  As explained above, the discretized problem is a
matrix eigenvalue problem $B_3 \B.\Psi=0$, where $B_3$ is a large
sparse matrix, whose rank depends on the mesh of the discretization,
and $\B.\Psi$ is the discretized version of the zero-mode $f$.  We
used NAG's sparse Gaussian elimination routines to find the zeros of
$\det(B_3)$, and determined the values of $\zeta_3$ for these zeros as
a function of $\zeta_h$.  The results of this procedure for space
dimensions $d=2,3,4$ are presented in Figs. 1,2, and 3. The zero modes
that correspond to any given value of $\zeta_3$ can be found
straightforwardly by inverse iterations of the matrix $B_3$ with an
arbitrary initial vector.

\subsection{Results}
The various branches shown in Figs. 1--3 can be organized on the basis
of the perturbation theory near $\zeta_h=0$ which was presented in
Section 2. At $\zeta_h=0$ we identify

\begin{figure}
\epsfxsize=8truecm
\label{Fig2}
\caption{ Same as Fig.1, but for $d=3$.}
\end{figure}
\begin{figure}
\epsfxsize=8truecm
\label{Fig3}
\caption{ Same as Fig.1, but for $d=4$. }
\end{figure}
\noindent
 the actual starting points of the branches with the analytic
 solutions for the lowest lying positive values of $\zeta_3$, which
 are $4,6,8$ etc.  In addition, for $d=2$ we observe the highest
 negative value which is $-2$. Measuring the slopes of the various
 branches at $\zeta_h=0$ we find full agreement with the perturbative
 predictions.

We see that in all dimensions the branch which begins at
$\zeta_h=0,\,\zeta_3=4$ continues in the non-perturbative region
without crossing any other branch until it ends at
$\zeta_h=2,\,\zeta_3=0$.  This branch is a continuation of the lowest
lying positive branch predicted by the perturbation theory.  The
negative branch (shown only for $d=2$) never rises above its
perturbative limit and is not relevant for the scaling behavior at any
value of $\zeta_h$. Note also that the point $\zeta_h=2,\, \zeta_3=0$
is an accumulation point of many branches, and we display only a part
of the actual spectrum near this point. These branches seem to
approach the accumulation point with a slope that grows without
limit. This finding is in agreement with the analytic result of the
perturbation calculation presented in Section 4.
\section{Conclusions}
It is well known by now that there exists a disagreement between the
scaling exponents $\zeta_4$ and the higher order exponents $\zeta_n$
computed via perturbative approaches and the predictions of another
approach based on the fully fused structure functions. The latter
approach seems to be consistent with the results of numerical
simulations in two spatial dimensions.  The main conclusion of the
present paper is that this disagreement cannot be ascribed to a formal
failure of the perturbation theory. If we accept the statement that
the scalar diffusivity is irrelevant, and compare the predictions of
perturbation theory at both ends of the range of the allowed values of
the parameter $\zeta_h$, we find that they are in excellent agreement
with the non-perturbative calculation of the scaling exponent, again
subject to the assumption that the diffusivity is
irrelevant. Therefore, if we want to understand the discrepancy
between the two approaches mentioned above, there are a few
possibilities that have to be sorted out by further research: \\ 
(i)
The crucial assumption that goes to the fully fused approach, which is
the linearity of the conditional average of the Laplacian of the
scalar, is wrong. \\ 
(ii) The computation of the zero modes which is achieved by discarding
the viscous terms in $\hat B_n$ is irrelevant for the physical
solution. It is not impossible that the limits $\zeta_h\to 0$ and
$\kappa \to 0$ do not commute, giving rise to some wicked properties
of the very small $\zeta_h$ regime.  That this is a possibility is
underlined by recent calculations of a shell model of the Kraichnan
model \cite{96WB}, in which it was shown that the addition of any
minute diffusivity changes the nature of the zero modes
qualitatively.\\ (iii) Lastly, and maybe most interestingly, it is
possible that the physical solution is not strictly scale invariant
through all the range of allowed distances\cite{Bob}. In other words,
it is possible that ${\cal F}_3(\B.r_1, \B.r_2,\B.r_3)$ is not a
homogeneous functions with a fixed homogeneity exponent $\zeta_3$, but
rather (for example), that $\zeta_3$ depends on the ratios of the
separations (or, in other words, the geometry of the triangle defined
by the coordinates). If this were also the case for even correlation
functions ${\cal F}_{2n}$, this would open an exciting route for
further research to understand how non-scale invariant correlation
functions turn, upon fusion, to scale invariant structure functions.

In light of the numerical results of Ref.~\cite{95KYC} and the
experimental results displayed in \cite{96CLP,96CLPP} we tend to doubt
option (i). If we were to guess at this point we would opt for
possibility (ii). More work however is needed to clarify this
important issue beyond doubt.
\appendix\section{The coefficients}
\begin{equation}
o_1:\quad
\begin{array}{rrrrrrcrrrrrr}
\{0,& 0,& 1,& -1,& 1,& 8\}&\quad&
\{0,& 0,& 2,& -2,& 0,& -2\}\\
\{0,& 0,& 2,& -1,& 1,& -4\}&\quad&
\{0,& 1,& 0,& -1,& 0,& -2\}\\
\{0,& 1,& 0,& 0,& 1,& -12\}&\quad&
\{0,& 1,& 0,& 2,& 1,& 8\}\\
\{0,& 2,& 0,& 0,& 0,& -2\}&\quad&
\{0,& 2,& 0,& 1,& 1,& -4\}\\
\{0,& 2,& 0,& 2,& 0,& 2\}&\quad&
\{0,& 2,& 0,& 3,& 1,& 4\}\\
\{1,& 0,& 1,& -1,& 1,& -8\}&\quad&
\{1,& 1,& 0,& 0,& 1,& 8\}\\
\{1,& 1,& 0,& 2,& 1,& -8\}&\quad&
\{2,& 0,& 0,& 0,& 0,& -2\}\\
\{2,& 0,& 0,& 1,& 1,& 4\}&&&&&&&
\end{array}
\end{equation}
\begin{equation}
o_2:\quad
\begin{array}{rrrrrrcrrrrrr}
\{0,& 0,& 1,& -1,& 1,& -4\}&\quad&
\{0,& 1,& 0,& 0,& 1,& 4\}\\
\{0,& 1,& 0,& 1,& 0,& 2\}&\quad&
\{1,& 0,& 0,& 0,& 0,& -2\}
\end{array}\end{equation}
\begin{equation}
o_3:\quad
\begin{array}{rrrrrrcrrrrrr}
\{0,& 0,& 1,& 0,& 2,& -4\}&\quad&
\{0,& 0,& 1,& -2,& 2,& 4\}\\
\{0,& 0,& 2,& 0,& 0,& 1\}&\quad&
\{0,& 0,& 2,& -2,& 0,& -1\}\\
\{0,& 0,& 2,& 0,& 2,& 2\}&\quad&
\{0,& 0,& 2,& -2,& 2,& -2\}\\
\{0,& 1,& 0,& -1,& 0,& -1\}&\quad&
\{0,& 1,& 0,& 1,& 0,& 3\}\\
\{0,& 1,& 0,& 3,& 0,& -2\}&\quad&
\{0,& 1,& 0,& 0,& 1,& -2\}\\
\{0,& 1,& 0,& 2,& 1,& 2\}&\quad&
\{0,& 1,& 0,& -1,& 2,& -2\}\\
\{0,& 1,& 0,& 1,& 2,& 2\}&\quad&
\{0,& 1,& 1,& 0,& 1,& 2\}\\
\{0,& 1,& 1,& 2,& 1,& -2\}&\quad&
\{0,& 1,& 1,& -1,& 2,& -4\}\\
\{0,& 1,& 1,& 1,& 2,& 4\}&\quad&
\{0,& 2,& 0,& 0,& 0,& -1\}\\
\{0,& 2,& 0,& 2,& 0,& 2\}&\quad&
\{0,& 2,& 0,& 4,& 0,& -1\}\\
\{0,& 2,& 0,& 1,& 1,& -2\}&\quad&
\{0,& 2,& 0,& 3,& 1,& 2\}\\
\{0,& 2,& 0,& 0,& 2,& 2\}&\quad&
\{0,& 2,& 0,& 2,& 2,& -2\}\\
\{1,& 0,& 1,& -1,& 1,& -2\}&\quad&
\{1,& 0,& 1,& 1,& 1,& 2\}\\
\{1,& 1,& 0,& 1,& 0,& -2\}&\quad&
\{1,& 1,& 0,& 3,& 0,& 2\}\\
\{1,& 1,& 0,& 0,& 1,& 2\}&\quad&
\{1,& 1,& 0,& 2,& 1,& -2\}\\
\{2,& 0,& 0,& 0,& 0,& 1\}&\quad&
\{2,& 0,& 0,& 2,& 0,& -1\}
\end{array}.
\end{equation}

{\bf Acknowledgments}: We thank Bob Kraichnan for useful suggestions
and J-P. Eckmann and Z. Olami for discussions. This work was supported
in part by the US-Israel BSF, the German-Israeli Foundation, the
Minerva Center for Nonlinear Physics and the Naftali and Anna
Backenroth-Bronicki Fund for Research in Chaos and Complexity.

\end{multicols}

\begin{references}
\bibitem{example} There are other possibilities that make ${\cal F}_3$
  relevant. One is putting on a fixed gradient, see \cite{95SS}, and
  another is to consider the fusion of three points in a higher order
  correlations, see \cite{96FGLP}.
\end{references}
\end{document}